\documentclass[amsmath,amssymb,aps,preprintnumbers,letterpaper,prx,twocolumn,longbibliography,superscriptaddress]{revtex4-1}

\usepackage{graphicx}
\usepackage{color}
\usepackage{enumitem}
\usepackage{picture}
\usepackage{calc}
\usepackage{float}
\usepackage{subfigure}
\usepackage{dsfont}
\usepackage{color}

\usepackage[utf8x]{inputenc}

\usepackage{tikz}
\usepackage[english]{babel}
\usepackage[T1]{fontenc}
\usepackage{indentfirst}
\usepackage{amsmath}
\usepackage{amssymb}
\usepackage{amsthm}
\usepackage{proof}
\usepackage{eufrak}
\usepackage{graphicx}
\usepackage{psfrag}
\usepackage{algorithm2e}

\usepackage[normalem]{ulem}

\allowhyphens

\newcommand{\tr}{\mathrm{Tr}\,}

\newcommand{\rd}{\mathrm{d}}
\newcommand{\re}{\mathrm{e}}

\newcommand{\ri}{\mathrm{i}}

\def\LT{\operatorname{LT}}


\usepackage{ifpdf}

\ifpdf
\usepackage{epstopdf}
\usepackage[pdftex,colorlinks,urlcolor=blue,citecolor=blue,linkcolor=blue]{hyperref}
\else
\usepackage[hypertex,colorlinks,urlcolor=blue,citecolor=blue,linkcolor=blue]{hyperref}
\fi
\begin{document}

\title{
  Exact Quench Dynamics from Algebraic Geometry
}

\author{Yunfeng Jiang}
\affiliation{Shing-Tung Yau Center and School of Physics, Southeast University, Nanjing 210096, China}
\author{Rui Wen}
\affiliation{Department of Physics and Astronomy, University of British Columbia,
6224 Agricultural Road, Vancouver, B.C. V6T 1Z1, Canada}
\author{Yang Zhang}
\affiliation{Interdisciplinary Center for Theoretical Study, University of Science and Technology of China,\\ Hefei,
Anhui 230026, China}
\affiliation{Peng Huanwu Center for Fundamental Theory, Hefei, Anhui 230026, China}

\preprint{USTC-ICTS/PCFT-21-34}

\begin{abstract}
  We develop a systematic approach to compute physical observables of integrable spin chains with finite length. Our method is based on Bethe ansatz solution of the integrable spin chain and computational algebraic geometry. The final results are analytic and no longer depend on Bethe roots. The computation is purely algebraic and does not rely on further assumptions or numerics. This method can be applied to compute a broad family of physical quantities in integrable quantum spin chains. We demonstrate the power of the method by computing two important quantities in quench dynamics: the diagonal entropy and the Loschmidt echo and obtain new analytic results.
\end{abstract}

\maketitle

\section{Introduction}
\label{sec:intro}
Finding exact solutions of important physical models has been a long-term endeavor. An important class of models that can be solved exactly are integrable models, such as the 2D Ising model and the Heisenberg spin chain. The exact solutions allow us to penetrate more deeply into the properties of the model and lead to new physical intuitions, which in many cases can be generalized to broader models including non-integrable ones.\par

A central question in integrable models is computing observables in a pure state $\langle\psi|\mathcal{O}_1\cdots\mathcal{O}_n|\psi\rangle$ or a thermal state $\text{tr}\left(\mathcal{O}_1\cdots\mathcal{O}_n\right)$ analytically. Here $\mathcal{O}_k$ are certain operators (not necessarily local). A common strategy for computing such quantities is inserting resolution of identity. For example,
\begin{align}
\label{eq:spectraldecomp}
\langle\psi|\mathcal{O}_1\mathcal{O}_2|\psi\rangle=&\,\sum_{\alpha}\langle\psi|\mathcal{O}_1|\alpha\rangle\langle\alpha|\mathcal{O}_2|\psi\rangle,\\\nonumber
\text{tr}\left(\mathcal{O}_1\mathcal{O}_2\right)=&\,\sum_{\alpha,\beta}\langle\alpha|\mathcal{O}_1|\beta\rangle\langle\beta|\mathcal{O}_2|\alpha\rangle,
\end{align}
where the sums on the right-hand side are over \emph{all} eigenstates of the Hamiltonian. To proceed further, one needs to find out quantities $\langle\psi|\mathcal{O}_i|\alpha\rangle$, $\langle\alpha|\mathcal{O}_i|\beta\rangle$ for all eigenstates and then perform the sum over states. For concreteness, let us now focus on integrable spin chains that can be solved by Bethe ansatz. For such models, the eigenstates $|\alpha\rangle$ can be constructed by Bethe ansatz and are parameterized by Bethe roots. Analytic results for the form factors $\langle\alpha|\mathcal{O}|\beta\rangle$ are known in many cases (see for example \cite{KorepinBook,MailletFF,SlavnovLecture}). On the other hand, performing the sum in (\ref{eq:spectraldecomp}) turns out to be a difficult task in general \footnote{For certain quantities such as the spin-spin correlation functions, such sums can be computed effectively by the master equation approach. See for example \cite{Kitanine:2004cp}.}. Typically one considers specific limits such as the thermodynamic limit where one of the states dominates and bypasses the need to compute the sum. For a finite spin chain, so far there is no systematic method to perform such sums.

In this paper, we develop a systematic method for computing such sums analytically. While we focus on the prototypical Heisenberg XXX spin chain in the current paper, we should emphasize that our method can be generalized to any Bethe ansatz solvable spin chains. Each eigenstate of XXX spin chain is labelled by a set of Bethe roots, which are physical solutions of the Bethe equations. Therefore, summing over all eigenstates is essentially equivalent to summing over all physical solutions of Bethe equations.
The crucial observation is, while finding a \emph{single} analytic solution of Bethe equation is hard or impossible, summing over \emph{all} solutions can be much simpler. We shall show that the sums in (\ref{eq:spectraldecomp}) can be performed analytically by using the proper tool --- computational algebraic geometry.\par

The algebro-geometric approach initiated in \cite{Jiang:2017phk} has been applied to computing partition functions of the 6-vertex model on a medium size lattice, both with torus \cite{LykkeJacobsen:2018nhn} and cylinder geometry \cite{Bajnok:2020xoz}. In this paper, we extent the method to a much wider class of observables. To demonstrate the power of our method, we consider two important quantities in quench dynamics, which are the diagonal R\'enyi entropy \cite{Barankov:2008qq,Alba:2017kdq,Alba2017Long} and the Loschmidt echo \cite{Gorin2006PhR,Jalabert2001PhRvL,Rossini2007PhRvA,Quan2006PhRvL,Pozsgay2013LE,Piroli:2016fpu,Piroli:2018amn}.


Before diving into technical details, let us first explain the main idea. The Loschmidt echo is defined by $\mathcal{L}_L(t)=|\mathcal{M}_L(\ri t)|^2$ where $\mathcal{M}_L(\omega)$ is given by
\begin{align}
\label{eq:Loschmidt}
\mathcal{M}_{L}(\omega)=\langle\psi|\re^{-\omega H}|\psi\rangle=\sum_{\alpha}|\langle\psi|\alpha\rangle|^2 \re^{-\omega E_{\alpha}}.
\end{align}

To evaluate (\ref{eq:Loschmidt}), naively we need to first find \emph{all} physical solutions of Bethe equations and then perform the sum. It is precisely at this point that computational algebraic geometry can play an important role. The upshot is, we do not need to solve the Bethe equation. Instead, we construct the \emph{companion matrices} for quantities like $\langle\psi|\alpha\rangle$ and $E_{\alpha}$. Performing the sum over states amounts to taking the trace of the companion matrix. The companion matrices are finite dimensional matrices, which can be constructed purely algebraically using Gr\"obner basis. Using this method, we obtain the following result
\begin{align}
\label{eq:ratioformM}
\mathcal{M}_{L}(\omega)=\frac{1}{2\pi i}\oint_{\mathcal{C}}F_L(z)\,e^{-\omega z}\rd z\,,
\end{align}
where $F_L(z)$ are \emph{rational functions} and the integration contour $\mathcal{C}$ is encircling all the poles counterclockwise. An explicit example will be given in section~\ref{sec:quench}. These rational functions can be worked out easily up to $L=20$ on a laptop. Compared to the original form (\ref{eq:Loschmidt}), the integral representation (\ref{eq:ratioformM}) is much more explicit and no longer depends on Bethe roots. The highly non-trivial job of solving Bethe equations and summing over all solutions have been fully accomplished.\par

The rest of the paper is structured as follows. In section~\ref{sec:XXXchain}, we give a brief review of Bethe ansatz of XXX spin chain. The main emphasis is on the completeness of Bethe ansatz. Section~\ref{sec:AG} introduces basic notions of computational algebraic geometry. In section~\ref{sec:quench}, we compute the diagonal entropy and Loschmidt echo using the algebro-geometric approach. We conclude in section~\ref{sec:concl} and discuss future directions.

\section{Bethe ansatz}
\label{sec:XXXchain}
In this section, we briefly review Bethe ansatz of XXX spin chain, with a special emphasis on completeness of Bethe ansatz.
The Heisenberg XXX spin chain is described by the following Hamiltonian
\begin{align}
\label{eq:HHeisenberg}
H=\frac{J}{4}\sum_{n=1}^L\left(\sigma_n^x\sigma_{n+1}^x+\sigma_n^y\sigma_{n+1}^y+\sigma_n^z\sigma_{n+1}^z-1\right).
\end{align}
We consider the periodic boundary condition $\sigma^\alpha_{L+1}=\sigma_1^{\alpha}$.

\paragraph{Primary and descendant states}
Eigenstates of (\ref{eq:HHeisenberg}) can be constructed by Bethe ansatz. An $N$-magnon ($N=1,2,\ldots,L$) state is characterized by $N$ Bethe roots $\mathbf{u}_N=\{u_1,\cdots,u_N\}$. We denote the corresponding eigenstate by $|\mathbf{u}_N\rangle$. The Bethe roots are physical solutions \footnote{It is well-known that some solutions of Bethe equations do not lead to eigenstates of the XXX Hamiltonian, such solutions are called non-physical.} of Bethe equations
\begin{align}
\label{eq:BAEor}
\left(\frac{u_j+\tfrac{i}{2}}{u_j-\tfrac{i}{2}}\right)^L=\prod_{k\ne j}^N\frac{u_j-u_k+i}{u_j-u_k-i}.
\end{align}
We distinguish two types of eigenstates, which are the \emph{primary} and \emph{descendant} states. The primary states have only \emph{finite} Bethe roots while descendant states contain roots at infinity.
Given a primary state $|\mathbf{u}_N\rangle$, its descendant states are obtained by acting operator $S^-$,
where
\begin{align}
S^{\pm}=\sum_{n=1}^L s_n^{\pm},\qquad s_n^{\pm}=\frac{1}{2}(\sigma_n^x\pm i\sigma_n^y).
\end{align}
In what follows, we will denote the descendant states by
\begin{align}
(S^-)^n|\mathbf{u}_N\rangle\equiv|\mathbf{u}_N,\infty^n\rangle.
\end{align}
Acting an $S^-$ operator on a Bethe state amounts to adding a Bethe root at infinity. The primary state $|\mathbf{u}_N\rangle$ and all its descendant states $|\mathbf{u}_N,\infty^n\rangle$ have the same energy, given by
\begin{align}
\label{eq:energy}
E(\mathbf{u}_N)=-\frac{J}{2}\sum_{k=1}^N\frac{1}{u_k^2+\tfrac{1}{4}}.
\end{align}
It is important to take into account both primary and descendant states when summing over states.

\paragraph{Completeness of Bethe ansatz}
An important question about Bethe ansatz concerns its completeness: does Bethe ansatz give \emph{all} eigenstates of the Hamiltonian ? This is a subtle question for generic integrable models. Fortunately, for XXX spin chain, the completeness problem has been studied extensively in the literature, see for example \cite{Baxter:2001sx,Mukhin2007,Hao:2013jqa}. The main conclusions are:
\emph{(i)} The number of \emph{physical solutions} of the Bethe equation (\ref{eq:BAEor})
is $\mathcal{N}_{L,N}^{\text{phys}}={L\choose N}-{L\choose N-1}$
for $N\le [L/2]$, where $[L/2]$ is the integer part of $[L/2]$.
\emph{(ii)} Eigenstates with $N>[L/2]$ corresponds to dual solutions of the Bethe equation. They can be obtained from the ones with $N\le[L/2]$ by flipping all the spins. From these, it is easy to verify that \emph{(iii)} Bethe ansatz is complete.

\paragraph{Rational $Q$-system}
The original form of the Bethe equations (\ref{eq:BAEor}) have solutions that are not physical. To select only the physical solutions, one can impose additional constraints, or more elegantly, reformulate Bethe equations. One nice reformulation of such kind is the rational $Q$-system \cite{Marboe:2016yyn,Bajnok:2019zub,Granet:2019knz}. For XXX spin chain, this is an alternative incarnation of the Wronskian relation of the Baxter's $TQ$-relation. The rational $Q$-system gives only physical solutions of Bethe equation and is easier to solve. Therefore in the algebro-geometric computations we will work with the rational $Q$-system instead of the Bethe equations. For more details and examples of rational $Q$-system, we refer to \cite{LykkeJacobsen:2018nhn,marboe2017ads}.\par

The rational $Q$-system gives a set of algebraic equations for the coefficients of the $Q$-function, which is defined by
\begin{align}
Q(u)=\prod_{j=1}^N(u-u_j)=u^N+\sum_{k=0}^{N-1}(-1)^k s_k u^k
\end{align}
where $\{u_1,\ldots,u_N\}$ are Bethe roots. The rational $Q$-system leads to a set of algebraic equations for $\{s_0,s_1,\ldots,s_{N-1}\}$.


\section{Algebraic geometry}
\label{sec:AG}
In this section, we introduce basic notions of computational algebraic
geometry which we need in what follows. A detailed and pedagogical
introduction to these notions in the context of Bethe ansatz can be
found in \cite{Jiang:2017phk}. Here we only highlight the main ideas.

Computational algebraic geometry is a modern tool to deal with
complicated algebraic varieties \cite{MR3330490}, with a broad application in
physics. We introduce this tool by posing a concrete question, which
will be solved by computational algebraic geometry.

Consider a set of $N$-variable algebraic equations
\begin{align}
\label{eq:FFeq}
F_1(z_1,\cdots,z_N)=\ldots =F_n(z_1,\cdots,z_N)=0.
\end{align}
We assume there are $\mathcal{N}$ solutions. We consider another polynomial $P(z_1,\cdots,z_N)$ and want to compute the following sum analytically
\begin{align}
\label{eq:defImulti}
\mathcal S[P]\equiv \sum_{\text{sol}}P(z_1,\cdots,z_N),
\end{align}
where we sum over the solutions of (\ref{eq:FFeq}). A brute force computation of this sum by numeric solutions is cumbersome
and suffers from numeric errors. Here we introduce the ingredients of
applying computational algebraic geometry to get the sum analytically.

\paragraph{Gr\"obner basis} The key point for this computation is to
reduce the given polynomial towards the equations (\ref{eq:FFeq}). However,  the
remainder is not unique. To have a well-defined remainder, we need to
transform the equations to a \emph{Gr\"obner basis}. Let
$I=\langle F_1\ldots F_n \rangle$ be the \emph{ideal} generated by $F_1,\ldots F_n$. The Gr\"obner basis is another set of polynomials $G_1,\cdots,G_m$ such that
$I=\langle G_1\ldots G_m \rangle$, with the additional
property that the remainder of the polynomial reduction for any
polynomial $P$ is unique with respect to
$\{G_j\}$. Gr\"obner basis can be understood as a nonlinear analogue of
Gaussian elimination and is computed by standard algorithms
\cite{MR3330490, DGPS}, which we review in the Supplemental Material. \par

\paragraph{Quotient ring} On the solution set of an equation system,
the value of a test function $P$ is well-defined modulo the ideal
$I$. Therefore we consider the quotient ring $A=\mathbb Q[z_1,\ldots
z_N]/I$. Over $\mathbb Q$, $A$ is a finite dimensional linear space
with the dimension
\begin{equation}
  \label{eq:1}
  \dim_{\mathbb Q} A= \mathcal N.
\end{equation}
The linear basis of $A$ is naturally the list of all monomials
$\{m_1\ldots m_\mathcal N\}$ which are
not divided by any leading terms of $\{ G_1\ldots G_m \}$.


\paragraph{Companion matrix} The values of a polynomial $P$ on the solution set can be
represented as the \emph{companion
  matrix} $\mathbf{M}_{P}$. In the quotient ring $A$, by the
polynomial division towards the Gr\"obner basis, we have
\begin{equation}
  \label{eq:3}
  P\,m_i = \sum_{j=1}^{\mathcal N} a_{ij} m_j, \quad a_{ij} \in \mathbb Q.
\end{equation}
We denote $\mathbf{M}_{P}$ as an $\mathcal N\times \mathcal N$
matrix with the entries $a_{ij}$'s. By the evaluation of the equation
above, the value of $P$ on a solution corresponds to an eigenvalue of
$\mathbf{M}_{P}$. Therefore we have the central formula for our purpose,
\begin{align}
\label{eq:trace_formula}
\mathcal S[P]=\tr\mathbf{M}_{P}.
\end{align}

Consider two polynomials $P_1$ and $P_2$ and their corresponding
companion matrices $\mathbf{M}_{P_1}$ and $\mathbf{M}_{P_2}$. The
companion matrix satisfies the following properties
\begin{align}
\label{eq:homomorph}
\mathbf{M}_{P_1\pm P_2}=&\,\mathbf{M}_{P_1}\pm\mathbf{M}_{P_2},\\\nonumber
\mathbf{M}_{P_1 P_2}=&\,\mathbf{M}_{P_1}\cdot\mathbf{M}_{P_2},\\\nonumber
\mathbf{M}_{P_1/P_2}=&\,\mathbf{M}_{P_1}\cdot\mathbf{M}_{P_2}^{-1}.
\end{align}
In particular, using the last property in (\ref{eq:homomorph}), we can
generalize the sum computation (\ref{eq:trace_formula}) for a rational function,
\begin{align}
\label{eq:rationalFunc}
\mathcal S[P_1/P_2]=\tr\left(\mathbf{M}_{P_1}\cdot\mathbf{M}_{P_2}^{-1}\right).
\end{align}
The construction of the Gr\"obner basis and the
companion matrix is purely arithmetic, and does not involve
algebraic extension or solving polynomial equations.

The method discussed here has been applied to the computation of partition functions of the 6-vertex model \cite{LykkeJacobsen:2018nhn,Bajnok:2020xoz}. In those cases, the function $P$ in (\ref{eq:defImulti}) is the eigenvalue of the transfer matrix $T(u_1,\cdots,u_N)$, which is a \emph{rational function} of the Bethe roots. Therefore the sum over solutions can be computed using (\ref{eq:rationalFunc}).\par

\paragraph{Beyond rational functions} In more general situations, we cannot restrict $P(u_1,\cdots,u_N)$ to be rational functions. One simple example is the thermal partition function of the XXX spin chain
\begin{align}
\label{eq:partZ}
Z_L(\beta)=\text{tr}\,e^{-\beta H}=\sum_{\alpha}\langle \alpha|e^{-\beta H}|\alpha\rangle,
\end{align}
which can be computed by Bethe ansatz. We first decompose the Hilbert space into different sectors of fixed magnon numbers. In each sector, the calculation boils down to computing the sum of the following type
\begin{align}
\sum_{\text{sol}}\re^{-\beta E_N(\mathbf{u})}
\end{align}
where the sum is over all physical solutions of Bethe equation. From (\ref{eq:energy}), it is clear that $e^{-\beta E_N(\mathbf{u})}$ is not a rational function of $\mathbf{u}$ and our method does not apply directly.
To write down an explicit analytic result, we seek for an alternative representation of $Z_L(\beta)$.\par

{To incorporate this more general situation into our method, let us consider the following sum
\begin{align}
\label{eq:genSum}
\sum_{\text{sol}}p(\mathbf{u})F(q(\mathbf{u}))
\end{align}
where $p(\mathbf{u})$ and $q(\mathbf{u})$ are rational functions of $\{u_1,\cdots,u_N\}$ and $F(z)$ can be any function that do not have singularities at $z=q(\mathbf{u})$. We can rewrite the sum (\ref{eq:genSum}) as
\begin{align}
\sum_{\text{sol}}p(\mathbf{u})F(q(\mathbf{u}))=\oint_{\mathcal{C}}\frac{\rd z}{2\pi i}\,F(z)\sum_{\text{sol}}\frac{p(\mathbf{u})}{z-q(\mathbf{u})},
\end{align}
where the contour encircles all possible values of $q(\mathbf{u})$. Denoting the companion matrices of $p(\mathbf{u})$ and $q(\mathbf{u})$ by $\mathbf{M}_p$ and $\mathbf{M}_q$ respectively, we have
\begin{align}
\label{eq:contourRep}
\sum_{\text{sol}}p(\mathbf{u})F(q(\mathbf{u}))=\oint_{\mathcal{C}}\frac{\rd z}{2\pi i}\,F(z)\,\text{tr}\,\left[\mathbf{M}_p(z-\mathbf{M}_q)^{-1}\right].
\end{align}
It is straightforward to write down similar contour integral representations for more complicated sums.}
\section{Exact quench dynamics}
\label{sec:quench}
In this section, we compute two important quantities in quench dynamics as concrete examples of our general method outlined in the previous sections. 
 The two quantities are defined in the following.

\paragraph{Diagonal R\'enyi entropy} We define the diagonal ensemble by the following density matrix
\begin{align}
\rho_d=\sum_{m}O_m|m\rangle\langle m|,
\end{align}
where the states $|m\rangle$ are eigenstates of the XXX spin chain and the coefficients $O_m$ are the overlaps with the initial state
\begin{align}
\label{eq:wm}
O_m=|\langle\Psi_0|m\rangle|^2.
\end{align}
The diagonal R\'enyi entropy is defined by
\begin{align}
    \label{eq:de}
S_d^{(\alpha)}\equiv\frac{1}{1-\alpha}\log\tr\rho_d^{\alpha}=\frac{1}{1-\alpha}\log\sum_m O_m^{\alpha}.
\end{align}

For fixed and integer values of $\alpha$, the overlap in (\ref{eq:de}) is a rational function of rapidities and the sum can be calculated analytically .

\paragraph{Loschmidt echo} The Loschmidt amplitude has been defined in (\ref{eq:Loschmidt}) which we quote here
\begin{align}
\label{eq:defLoschmidt}
\mathcal{M}_L(\omega)=\langle\Psi_0|\re^{-\omega H}|\Psi_0\rangle=\sum_m O_m\,\re^{-\omega E_m}
\end{align}
where $O_m$ has been defined in (\ref{eq:wm}) and $E_m$ is the energy of state $|m\rangle$. The Loschmidt echo is given by $\mathcal{L}(t)=|\mathcal{M}_L(\ri t)|^2$.
Similar to the thermal partition function, the analytic result of the sum in the Loschmidt amplitude (\ref{eq:defLoschmidt}) cannot be written down directly. We will give the result in the contour integral representation (\ref{eq:contourRep}).

\paragraph{Integrable quench}
We consider the integrable quench where the initial state $|\Psi_0\rangle$ is an integrable initial state \cite{Piroli:2017sei}. Such states have a number of nice properties which make them especially suitable for analytical studies. In particular,
the overlap $O_m=|\langle\Psi_0|m\rangle|^2$, which is an important ingredient for both the diagonal entropy (\ref{eq:de}) and the Loschmidt echo (\ref{eq:defLoschmidt}) can be written down explicitly as a rational function of rapidities \cite{Brockmann,Pozsgay:2014,Foda:2015nfk,deLeeuw:2019ebw,Jiang:2020sdw}.\par

For a Bethe state with $N$ rapidities (assuming $N$ is even for simplicity), the overlap $\langle\Psi_0|\mathbf{u}_N\rangle$ is non-zero only if the Bethe roots are paired, namely
\begin{align}
\label{eq:pair1}
\mathbf{u}_N=\{u_1,-u_1,u_2,-u_2,\ldots,u_{\frac{N}{2}},-u_{\frac{N}{2}}\}.
\end{align}
A simple example of an integrable initial state is the N\'eel state
\begin{align}
\label{eq:Neel}
|\Psi_0\rangle=\frac{1}{2}\left(|\!\uparrow\downarrow\rangle^{\otimes L/2}+|\!\downarrow\uparrow\rangle^{\otimes L/2}\right),
\end{align}
 Due to magnon number conservation, the states which have non-zero overlaps with N\'eel state (\ref{eq:Neel}) are $L/2$-magnon states, which consists of $|\mathbf{u}_N,\infty^{L/2-N}\rangle$ $(N=0,1,\cdots,L/2)$. The overlap is given by \cite{brockmann2014gaudin}
\small{
\begin{align}
\label{eq:overlapFormula}
\frac{\langle\Psi_0|\mathbf{u}_N\rangle}{\sqrt{\langle\mathbf{u}_N|\mathbf{u}_N\rangle}}=\frac{\sqrt{2}(L/2-N)!}{\sqrt{(L-2N)!}}
\prod_{j=1}^{N/2}\frac{\sqrt{u_j^2+\tfrac{1}{4}}}{4u_j}\sqrt{\frac{\det G^+}{\det G^-}}
\end{align}
}
where $|\mathbf{u}_N\rangle\equiv|\mathbf{u}_N,\infty^{L/2-N}\rangle$ and $G^{\pm}$ is the Gaudin matrix
\begin{align}
&G_{jk}^{\pm}=\delta_{jk}\left(L K_{1/2}(u_j)-\sum_{l=1}^{M/2}K_1^+(u_j,u_l)\right)+K_1^{\pm}(u_j,u_k),\\\nonumber
&K_{\alpha}^{\pm}(u,v)=K_{\alpha}(u-v)\pm K_{\alpha}(u+v),\\\nonumber
&K_{\alpha}(u)=\frac{2\alpha}{x^2+\alpha^2}
\end{align}

\paragraph{Diagonal R\'enyi entropy}
For the computation of diagonal R\'enyi, the crucial quantity is the sum $\sum_m O_m^{\alpha}$.
For a N\'eel state $|\Psi_0\rangle$, the nonzero overlaps come from Bethe states $|\mathbf{u}_N,\infty^{L/2-N}\rangle$ with $N=0,2,\cdots,2\lfloor L/4\rfloor$. If $\alpha$ is an integer, $O_m^{\alpha}$ is a rational function of Bethe roots. The sum can thus be performed by algebro-geometric method directly. We denote the companion matrix of
\begin{align}
\label{eq:defOM}
O_N(\mathbf{u}_N)=\frac{\left|\langle\mathbf{u}_N,\infty^{L/2-N}|\Psi_0\rangle\right|^2}{\langle\mathbf{u}_N,\infty^{L/2-N}|\mathbf{u}_N,\infty^{L/2-N}\rangle}
\end{align}
by $\mathbf{M}_{O_N}$. Using the property (\ref{eq:homomorph}), the diagonal R\'enyi entropy can be written as
\begin{align}
\label{eq:Sdag}
    S_d^{\alpha}(L)=\frac{1}{1-\alpha}\log \sum_{N=0}^{2[ L/4]}\tr(\mathbf{M}_{O_N})^{\alpha}
\end{align}
We study an explicit example for $L=8$ as an illustration. In this case only Bethe states $|\mathbf{u}_N,\infty^{4-N}\rangle$ with $N=0,2,4$ contribute. We discuss the $N=4$ sector in detail. The other sectors are simpler and can be computed in a similar way.\par
In the $M=4$ sector, the paired Bethe roots take the form $\mathbf{u}_4=\{u_1,-u_1,u_2,-u_2\}$. The overlap square $O_4(u_1,u_2)$ is a symmetric rational function in the rapidities $u_1,u_2$.  The ideal consists of the algebraic equations from rational $Q$-system (see Supplemental Material for more details), together with the nonsingular condition \footnote{Bethe equations contain physical solutions of the form $\{\ri/2,-\ri/2,u_1,\cdots,u_{N-2}\}$ which are called singular solutions. The Bethe states correspond to such solutions have zero overlap with the N\'eel state. Therefore we can impose the nonsingular solution and exclude such solutions from the beginning.}
\begin{align}
\label{eq:nonsingular}
    w(u_1^2+1)(u_2^2+1)+1=0.
\end{align}
 Following the standard algorithm, we can compute a Gr\"obner basis of this ideal and construct the companion matrix of $O_4$, which reads
\begin{align}
    \mathbf{M}_{O_4}=
\begin{bmatrix}
 \frac{19}{156} & \frac{3}{13} & -\frac{109}{1872} \\[0.45em]
 -\frac{7}{6240} & \frac{19}{1560} & \frac{9}{8320} \\[0.45em]
 -\frac{5}{26} & -\frac{14}{13} & \frac{83}{312} \\
\end{bmatrix}.
\end{align}
The companion matrices of $O_2,O_0$ can be computed in a similar fashion. 
Then diagonal R\'enyi entropy can be computed straightforwardly. We list results for $\alpha=2,6,10$
\begin{align}
\label{eq:samplediagRE}
    &S_d^2(8)=\log \left(\frac{143325}{49009}\right),\\\nonumber
    &S_d^6(8)=\frac{1}{5} \log \left(\frac{38274471591890625}{512161566111913}\right),\\\nonumber
    &S_d^{10}(8)=\frac{1}{9} \log \left(\frac{112319474922585645380859375}{75385210067492164108951}\right).
\end{align}    \par
For any non-integer $\alpha>0$, using the representation(\ref{eq:contourRep}) we have 
\begin{widetext}
\begin{align}
  \label{eq:contourRe}
S_d^{\alpha}=&\frac{1}{1-\alpha}\log\left(\oint_{\mathcal{C}}\frac{\rd z}{2\pi i}\,z^{\alpha}\sum_{N=0}^{2\lfloor L/4\rfloor}\tr(z-\mathbf{M}_{O_N})^{-1})\right)\\\nonumber
=&\frac{1}{1-\alpha}\log\left(\oint_{\mathcal{C}}\frac{\rd z }{2\pi i}\,z^{\alpha}\left(\frac{21 \left(1323 z^2-504 z+20\right)}{9261 z^3-5292 z^2+420 z-8}+\frac{39 \left(1755 z^2-468 z+16\right)}{22815 z^3-9126 z^2+624 z-8}+\frac{35}{35z-1}\right)\right).
\end{align}
\end{widetext}
The diagonal R\'enyi entropy have been computed in the thermodynamic limit in \cite{Alba:2017kdq,Alba2017Long} using the quench action approach. Here we offer a general method for finite length spin chains which gives exact analytic results. Our approach is new and complimentary to the works \cite{Alba:2017kdq,Alba2017Long} in the thermodynamic limit.
%
%

\paragraph{Loschmidt echo}
We apply the contour integral representation to compute the Loschmidt echo.
The Loschmidt amplitude can be written as
\begin{align}
  \mathcal{M}_L(\omega)
  =\oint_{\mathcal{C}}\frac{\rd z}{2\pi i}\,\widetilde{\mathcal{M}}_L(z)\,\re^{-\omega z},
  \end{align}
where
\begin{align}
\label{eq:companionME}
\widetilde{\mathcal{M}}_L(z)&:=\sum_{k=0}^{[L/4]}\widetilde{\mathcal{M}}_L^{(2[ L/4]-2k)}(z),\\\nonumber
\widetilde{\mathcal{M}}_L^{(N)}(z)&:=\tr\left(\mathbf{M}_{O_N}\cdot(z\mathbf{1}-\mathbf{M}_{E_N})^{-1}\right).
\end{align}
Apart from the overlap matrix $\mathbf{M}_{O_N}$, we also need the companion matrix of the energy $E_N=E(\mathbf{u}_N)$.
For the $L=8$ example, only the states with $N=4,2,0$ contribute. In the $N=4$ sector, the energy is (taking $J=1$) given by
\begin{align}
    E_4=-\frac{16 s_1+8}{4 s_1+16 s_3+1}.
\end{align}
The companion matrix can reads
\begin{align}
\mathbf{M}_{E_4}=
    \begin{bmatrix}
     -\frac{27}{8} & -\frac{5}{2} & \frac{65}{96} \\[0.4em]
     \frac{1}{64} & -\frac{25}{16} & -\frac{11}{768} \\[0.4em]
     \frac{9}{4} & 15 & -\frac{81}{16} \\
    \end{bmatrix}.
\end{align}
Using (\ref{eq:companionME}), we obtain
\begin{align}
\label{eq:LE84}
    \widetilde{\mathcal{M}}^{(4)}_8(z)=\frac{2 \left(3 z^2+15 z+17\right)}{15 \left(z^3+10 z^2+29 z+25\right)}.
\end{align}
The contributions of $M=0,2$ can be calculated in the same way. Combining them, we get
\begin{align}
\label{eq:anaLs8}
    \widetilde{\mathcal{M}}_8(z)=\frac{z^6+13 z^5+63 z^4+143 z^3+153 z^2+65 z+5}{z \left(z^3+10 z^2+29 z+25\right) \left(z^3+7 z^2+14 z+7\right)}.
\end{align}
The Loschmidt amplitude then takes the contour integral form
\begin{align}
\label{eq:anaLyticL}
\mathcal{M}_8(\omega)=\oint_{\mathcal{C}}\frac{\rd z}{2\pi i}\widetilde{\mathcal{M}}_8(z)e^{-\omega z}
\end{align}
The integral can be evaluated by various methods straightforwardly. We give more results in the Supplemental Material. \par
In \cite{Piroli:2016fpu,Piroli:2018amn},   the analytical results of Loschmidt echo is obtained in the thermodynamic limit using quantum transfer matrix method. Here we give a systematic method to compute it analytically for any finite length spin chain. To the best of our knowledge, the analytical result of the form (\ref{eq:anaLyticL}) is new.

\section{Conclusions}
\label{sec:concl}
In this paper, we present a systematic method to compute a large family of physical observables for the finite length XXX spin chain. Our method is based on Bethe ansatz solution of the spin chain and computational algebraic geometry. We applied this method to compute the diagonal R\'enyi entropy and the Loschmidt echo, obtaining new analytic results for both quantities in (\ref{eq:samplediagRE}) and (\ref{eq:anaLs8}). Our method is completely algebraic and avoids the need of solving Bethe equations.
Apparently there are many interesting directions to explore based on our current results. For the XXX spin chain, it would be interesting to compute other important quantities such as correlation functions of local spin operators. These quantities are considerably more involved than the Loschmidt echo and diagonal entropy. To compute such quantities for relatively long spin chains $(L\sim 20)$, a more efficient implementation of our method is needed.
The general strategy in this paper clearly generalizes to other types of spin chains, such as the XXZ spin chain and chains solvable by the nested Bethe ansatz. Naively the Bethe equations of the XXZ spin chain involve hyperbolic functions and our approach, which works for polynomial functions do not apply directly. However, by a change of variable to the multiplicative variables, the Bethe equation can be brought to a polynomial form. The rational $Q$-system for the XXZ spin chain has been studied in \cite{Bajnok:2019zub}.
Finally, it is worth noting that the Gr\"obner basis is a general tool which works for any polynomial equations. It is therefore desirable to find alternative ways to construct the companion matrices, which is more fine-tuned for integrable models.

\vspace{0.1cm}
{\bf Acknowledgments}

We are grateful to Balazs Pozsgay and Jing Deng for useful comments on the draft and
discussions. YZ is supported from the NSF of China through Grant
No. 11947301, 12047502 and No. 12075234.

\bibliographystyle{apsrev4-2}
\bibliography{yunfeng}
\clearpage

\widetext
\begin{center}
\textbf{\large Supplemental Materials for `Exact Quench Dynamics from Algebraic Geometry'}
\end{center}
\begin{center}
Yunfeng Jiang, Rui Wen, Yang Zhang
\end{center}

\section{Basic algorithms in computational algebraic geometry}

In this appendix, we review some basic algorithms in computational algebraic
geometry \cite{MR3330490} used in this paper.

We define a monomial order $\succ$ as a total order of all monomials
of a polynomial ring $R$ with the following properties,
\begin{enumerate}
\item $\succ$ respects the product, \emph{i.e.}, if $u\succ v$ then for
 $uw\succ vw$.
\item $ u\succ 1$, if $u$ is not non-constant monomial.
  \end{enumerate}
With the monomial order $\succ$, the highest monomial of a polynomial $F$
is defined as the leading term of $F$, $\LT(F)$.

The multivariate polynomial division (Algorithm.\ref{Algorithm:MPD}) serves as the foundation of most
computational algebraic geometry algorithm. Given the input polynomial
$F$, the divsors $f_1,\ldots f_k$, and a monomial ordering $\succ$,
the division reads
\begin{equation}
  \label{eq:1}
  F= \sum_{i=1}^k q_i f_i +r
\end{equation}
where $r$ is the remainder. The leading term of $r$ does not divide
any leading term of $f_i$'s. We use the abbreviation $\overline
F^{f_1,\ldots f_k}$ for the remainder $r$. Note that unless $f_1,\ldots f_k$ form a
Gr\"obner basis, the remainder is not unique.
\begin{algorithm}[h]
 \KwData{$F$, $f_1\ldots f_k$ and a monomial order  $\succ$}
$q_1:=\ldots :=q_k=0$, $r:=0$\;
\While{$F! =0$}{
         $reductionstatus:=0$\;
         \For{$i:=1$ to $k$}{
          \If{$\LT(f_i)|\LT(F)$}{
                      $q_i:=q_i+\frac{\LT(F)}{\LT(f_i)}$\;
                      $F:=F-\frac{\LT(F)}{\LT(f_i)} f_i$\;
                      $reductionstatus:=1$\;
                      \textbf{break}
          }
           }
           \If{$reductionstatus=0$}{
                      $r:=r+\LT(F)$\;
                      $F:=F-\LT(F)$\;
           }
}
\Return  $q_1\ldots q_k$, $r$
\label{Algorithm:MPD}
 \caption{Multivariate polynomial division}
\end{algorithm}

To get a unique remainder, we need to transform the divisors
$f_1,\ldots f_k$ to a Groebner basis. The basic algorithm to compute
Gr\"obner basis is the Buchburger algorithm (Algorithm. \ref{Algorithm:Buchberger}).
\begin{algorithm}[h]
\KwData{$B=\{f_1\ldots f_n\}$ and a monomial order $\succ$}
$queue:=\text{all subsets of B with exactly two elements}$\;
\While{$queue!=\emptyset$}{
                     $\{f,g\}:=\text{head of } queue$\;
                      $r:=\overline{S(f,g)}^B$\;
                     \If{$r\not=0$}{
                                            $B:=B\cup{r}$\;
                                            {\bf enqueue} $\{\{B_1,r\},\ldots \{{\text{last\ of}\ }B,r\}\}$\;
                     }
                     {\bf dequeue}\;
}
\Return $B$ a Groebner basis
\label{Algorithm:Buchberger}
\caption{Buchberger algorithm}
\end{algorithm}

The Gr\"obner basis for polynomials is like the row reduced row echlon form for a matrix in
linear algebra, from which a lot of interesting computation can be
done. If the equation $f_1=\ldots f_k=0$ has $\mathcal N$ solutions, then
there are exactly $\mathcal N$
monomials, $m_1,\ldots m_{\mathcal
  N}$, which are not divided by any leading term of the
corresponding Groebner basis. Such monomials form the {\it linear
  basis} of the quotient ring $R/\langle f_1 ,\ldots
f_k\rangle$. The companion matrix of a polynomial $P$, can be constructed from the following
algorithm (Algorithm. \ref{Algorithm:Companion}):
\begin{algorithm}[h]
  \KwData{$P$, $m_1,\ldots m_{\mathcal
  N}$ a linear basis, $G$ a Gr\"obner basis}
  $M:=$ an $\mathcal N\times \mathcal N$ empty matrix\;
\For{$i:=1$ to $\mathcal N$}{
      $r:=\overline{m_i P}^G$\;
      {\bf expand} $r$ as the sum $\sum_{j=1}^{\mathcal N} a_j m_j$\;
     \For{$j:=1$ to $\mathcal N$}{
             $M_{ij}:=a_j$\;
       }
}
\Return $M$ the companion matrix of $P$
\label{Algorithm:Companion}
\caption{Companion Matrix}
\end{algorithm}

All the algorithm mentioned here are implemented in the computer
algebra system {\sc Singular} \cite{DGPS}.

\section{More on diagonal R\'enyi entropy}
In this section, we give more details and results on the computation of diagonal R\'enyi entropy.
\subsection{The $L=8$ example}
Let us first present more details for the $L=8$ example discussed in the main text. The length-8 N\'eel state is a 4-magnon state. The Bethe states with non-zero overlap with $|\Psi_0\rangle$ are of the form $|\mathbf{u}_N,\infty^{4-N}\rangle$, ($N=4,2,0$) with paired finite rapidities $\mathbf{u}_N:=\{u_1,-u_1,\cdots,u_{N/2},-u_{N/2}\}$.\par
For $N=4$, we have $\mathbf{u}_4=\{u_1,-u_1,u_2,-u_2\}$. To write down the ideal for this sector, we apply the rational $Q$-system approach \cite{Marboe:2016yyn}. It is more convenient to work with the following symmetric combinations of $u_1,u_2$
\begin{align}
s_1=u_1^2+u_2^2,\qquad s_2=u_1^2u_2^2.
\end{align}
The rational $Q$-system leads to the following set of algebraic equations for $s_1$ and $s_3$
    \begin{align}
    \label{eq:ideal8}
       &48 s_1^3+48 s_1^2-352 s_2 s_1-6 s_1-16 s_2-3=0,\\\nonumber
        &768 s_2 s_1^3+1536 s_2 s_1^2-3328 s_2^2 s_1+672 s_2 s_1-s_1+256 s_2^2-64 s_2-1=0,\\\nonumber
        &-384 s_1^4+768 s_1^3-2048 s_2 s_1^2+336 s_1^2-640 s_2 s_1-72 s_1+1280 s_2^2-224 s_2-27=0.
    \end{align}
The non-singular condition(\ref{eq:nonsingular}) can be expressed in terms of $s_1,s_2$ as:
\begin{align}
    \label{eq:nonsingular}
        w\left(\frac{s_1}{4}+s_2+\frac{1}{16}\right)+1=0.
    \end{align}
The equations (\ref{eq:ideal8}) and (\ref{eq:nonsingular}) give the ideal. We can compute a Gr\"obner basis of the ideal by Algorithm~\ref{Algorithm:Buchberger}. It can be done by the CAG software \textit{Singular} \cite{DGPS}.

The next step is computing the companion matrix of $O_4$, which can be written in terms of $s_1,s_3$ as
\begin{align}
        O_4=&\,\frac{|\langle\Psi_0|\mathbf{u}_4\rangle|^2}{\langle\mathbf{u}_4|\mathbf{u}_4\rangle}=\frac{8 s_1^3+32 s_2 s_1^2+22 s_1^2+16 s_2 s_1+17 s_1-256 s_2^2+32 s_2+3}{2688 s_2 s_1^2+7680 s_2 s_1-23040 s_2^2+4608 s_2}.
\end{align}
This can be done by Algorithm~\ref{Algorithm:Companion}. The result reads
\begin{align}
    \mathbf{M}_{O_4}=
\begin{bmatrix}
 \frac{19}{156} & \frac{3}{13} & -\frac{109}{1872} \\[0.45em]
 -\frac{7}{6240} & \frac{19}{1560} & \frac{9}{8320} \\[0.45em]
 -\frac{5}{26} & -\frac{14}{13} & \frac{83}{312} \\
\end{bmatrix}.
\end{align}
For $N=2$, we have $\mathbf{u}_2=\{u_1,-u_1\}$. We define $s_1=u_1^2$, which satisfies the following equation from rational $Q$-system
\begin{align}
    -1792 s_1^4-1792 s_1^3+224 s_1^2+80 s_1+1=0.
\end{align}
Together with the non-singular condition
\begin{align}
    w s_1+1=0,
\end{align}
give the ideal for the $N=2$ sector. The computation of the Gr\"obner basis is much simpler in this case.
The overlap for $N=2$ reads
\begin{align}
O_2=\frac{|\langle\Psi_0|\mathbf{u}_2,\infty^2\rangle|^2}{|\langle\mathbf{u}_2,\infty^2|\mathbf{u}_2,\infty^2\rangle|^2}=\frac{4 u_1^2+1}{168 u_1^2}=\frac{4s_1+1}{168s_1}.
\end{align}
The companion matrix is computed similarly
\begin{align}
    \mathbf{M}_{O_2}=
        \begin{bmatrix}
         \frac{1}{42} & 0 & -\frac{1}{168} \\[0.45em]
         \frac{8}{3} & \frac{32}{21} & 4 \\[0.45em]
         -\frac{2}{3} & -\frac{8}{21} & -\frac{41}{42} \\
        \end{bmatrix}.
\end{align}
Finally the overlap for $N=0$ is simply a constant, and we have $\mathbf{M}_{O_4}=\frac{1}{35}$. Using the formula
\begin{align}
    \label{eq:Sdag}
        S_d^{\alpha}(L)=\frac{1}{1-\alpha}\log \sum_{M=0}^{2[ L/4]}\tr(\mathbf{M}_{O_M})^{\alpha},
    \end{align}
one can get diagonal R\'enyi entropy for any integer value of $\alpha$ easily. We list a few more results in addition to the ones given in the main text
\begin{align}
    S^{12}_d(8)=&\frac{1}{11} \log \left(\frac{1464935175638442473803261962890625}{225358488016652052246678510097}\right),\\\nonumber
    S^{16}_d(8)=&\frac{1}{15} \log \left(\frac{477663172460063971194114484891204833984375}{3906838765549548127722792233690449591}\right),\\\nonumber
    S^{20}_d(8)=&\frac{1}{19} \log \left(\frac{8036178252665965191421547988006495102067279815673828125}{3510225873887152206208588744031993539782535336781}\right).
\end{align}

\subsection{Results for higher $L$}
It is straightforward to compute results for higher $L$ following the same approach. For example, the results for $S^2_d(L)$ for $L=10,12,14$ are given by
\begin{align*}
    S_{d}^2(10)&=\log \left(\frac{599841622044}{162613034999}\right), \\
    S_d^2(12)&=\log \left(\frac{3039576559583745895715487507208261594578135}{485206057903650436993894416560599575805337}\right),\\
    S_d^2(14)&=\log \left(\frac{166772292644315654600877189788011138526647981072451419242370373797233905494807019421605475}{19375258475703026886733750017199427784747776297026681635447971033890730345637366575186858}\right).
\end{align*}
We have computed the companion matrices up to $L=20$, which can be download from \textcolor{blue}{data link}. From the quench action approach, the authors of \cite{alba2017quench} found that for a given $\alpha$, the the R\'enyi entropy grows linearly with length $L$ as $L\to\infty$. We compared the exact result with their linear fit for $\alpha=2,3,4$, as is shown in figure~\ref{Fig.main2}.
\begin{figure}[H]
    \centering 
    \includegraphics[width=0.32\textwidth]{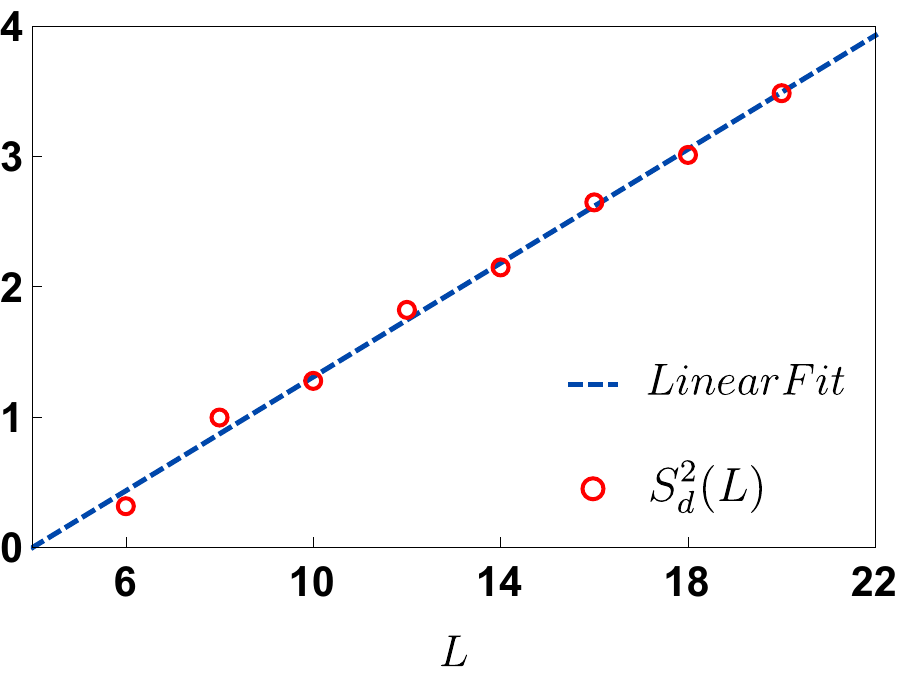} 
    \includegraphics[width=0.32\textwidth]{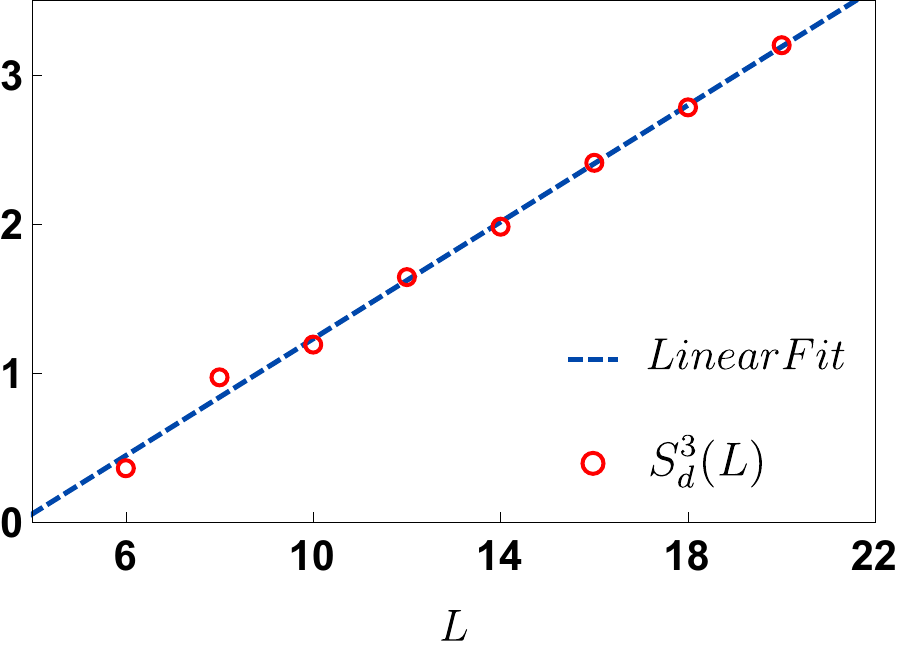}
    \includegraphics[width=0.32\textwidth]{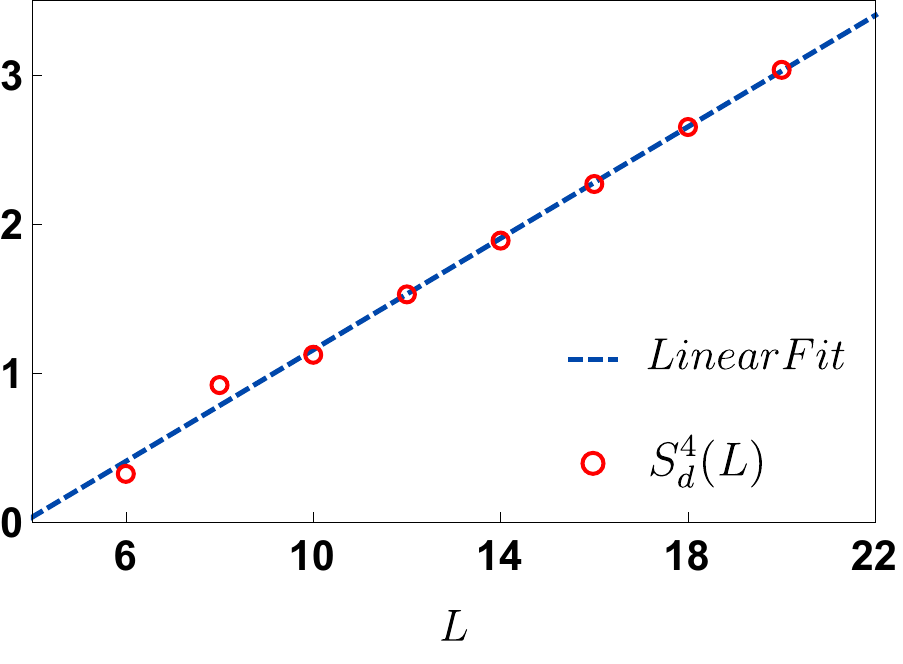}
    \caption{Diagonal R\'enyi entropy and their linear fit for $\alpha=2,3,4$. For each $\alpha$, we computed the data points for $L$ up to $20$.} 
    \label{Fig.main2} 
\end{figure}
As we can see, as $L$ grows, the data points fit well with their linear fit, which is consistent with the prediction \cite{alba2017quench}.

\section{More on Loschmidt echo}
The algebro-geometric computation of the Loschmidt echo is similar to the one of diagonal R\'enyi entropy. The main difference is that the final result is given in terms of a contour integral. To obtain the real time dynamics of Loschmidt echo, we need to compute the contour integral explicitly. This can only be done numerically, but it is quite straightforward. For example, in the $L=8$ case, the Loschmidt amplitude (\ref{eq:anaLyticL}) reads
\begin{align}
    \mathcal{M}_8(\omega)=\oint_{\mathcal{C}}\frac{\rd z}{2\pi i}\frac{z^6-13 z^5+63 z^4-143 z^3+153 z^2-65 z+5}{z \left(z^3-10 z^2+29 z-25\right) \left(z^3-7 z^2+14 z-7\right)}e^{-\omega z}.
\end{align}
The integral can be calculated by residue theorem. We give explicit numerical results for $L=8,10$:
\begin{align}
    \mathcal{L}_8(t)&=0.00006242 (1.000 \cos (0.4679 t)+0.8538 \cos (1.011 t)+1.505 \cos (1.653 t)+0.7397 \cos (1.830 t)\\\nonumber
    &+8.420 \cos (2.676 t)+3.532 \cos (3.000 t)+29.28 \cos (3.879 t)+2.994 \cos (3.952 t)+24.77 \cos (4.939 t)\\\nonumber
    &+53.05 \cos (6.592 t)+1.009) \cos (30.00 t) (52.57 \cos (23.41 t)+24.54 \cos (25.06 t)+2.966 \cos (26.05 t)\\\nonumber
    &+29.02 \cos (26.12 t)+3.500 \cos (27.00 t)+8.343 \cos (27.32 t)+0.7329 \cos (28.17 t)+1.491 \cos (28.35 t)\\\nonumber
    &+0.8460 \cos (28.99 t)+0.9909 \cos (29.53 t)+1.000 \cos (30.00 t))
\end{align}
\begin{align}
    \mathcal{L}_{10}(t)&=4.580\times 10^{-6} (1.023+1.000 \cos (0.3175 t)+0.9078 \cos (0.6812 t)+0.6344 \cos (1.112 t)\\\nonumber
    &+1.301 \cos (1.169 t)+1.106 \cos (1.369 t)+0.01275 \cos (1.561 t)+0.5705 \cos (1.906 t)+2.147 \cos (2.285 t)\\\nonumber
    &+2.783 \cos (2.444 t)+16.52 \cos (2.668 t)+1.919 \cos (2.818 t)+0.6312 \cos (3.147 t)+5.335 \cos (3.310 t)\\\nonumber
    &+5.267 \cos (3.789 t)+4.769 \cos (3.916 t)+45.46 \cos (3.919 t)+40.64 \cos (4.571 t)+2.191 \cos (4.614 t)\\\nonumber
    &+7.210 \cos (5.154 t)+7.099 \cos (5.204 t)+18.98 \cos (5.402 t)+61.09 \cos (5.847 t)+24.93 \cos (6.340 t)\\\nonumber
    &+146.6 \cos (7.071 t)+72.50 \cos (8.387 t)+1.023) \cos (89.00 t) (70.88 \cos (80.61 t)+143.3 \cos (81.93 t)\\\nonumber
    &+24.37 \cos (82.66 t)+59.73 \cos (83.15 t)+18.56 \cos (83.60 t)+6.940 \cos (83.80 t)+7.048 \cos (83.85 t)\\\nonumber
    &+2.142 \cos (84.39 t)+39.73 \cos (84.43 t)+49.10 \cos (85.08 t)+5.149 \cos (85.21 t)+5.215 \cos (85.69 t)\\\nonumber
    &+0.6170 \cos (85.85 t)+1.876 \cos (86.18 t)+16.15 \cos (86.33 t)+2.720 \cos (86.56 t)+2.099 \cos (86.72 t)\\\nonumber
    &+0.5577 \cos (87.09 t)+0.01247 \cos (87.44 t)+1.081 \cos (87.63 t)+1.272 \cos (87.83 t)+0.6202 \cos (87.89 t)\\\nonumber
    &+0.8875 \cos (88.32 t)+0.9776 \cos (88.68 t)+1.000 \cos (89.00 t))
\end{align}
Results for higher values of $L$ can be downloaded from \textcolor{blue}{link}.\par

The time evolution of Loschmidt echo for $L=16,18,20$ are given in figure~\ref{Fig.main2}. It can be seen that in all these cases the Loschmidt echo first decay quickly and then fluctuate around a mean value, which is predicted to be $\widetilde{\mathcal{L}}_L=\sum_mO_m^4$ \cite{Lorenzo2017Losecho}. The fluctuation decreases as $L$ increases. These behavior are in agreement with the predictions in \cite{Lorenzo2017Losecho}.
\begin{figure}[H] 
    \centering 
    \includegraphics[width=0.32\textwidth]{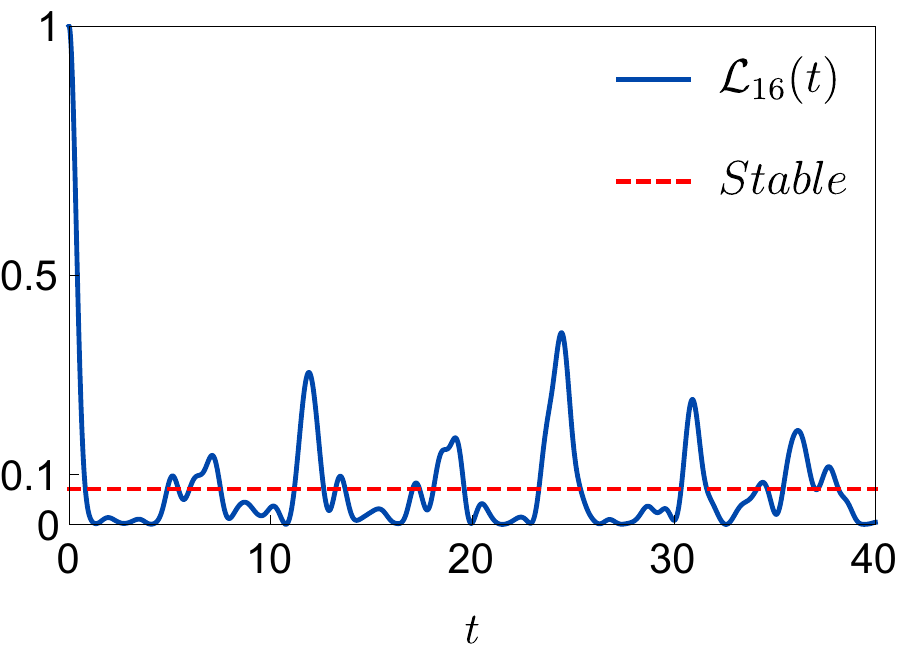} 
    \includegraphics[width=0.32\textwidth]{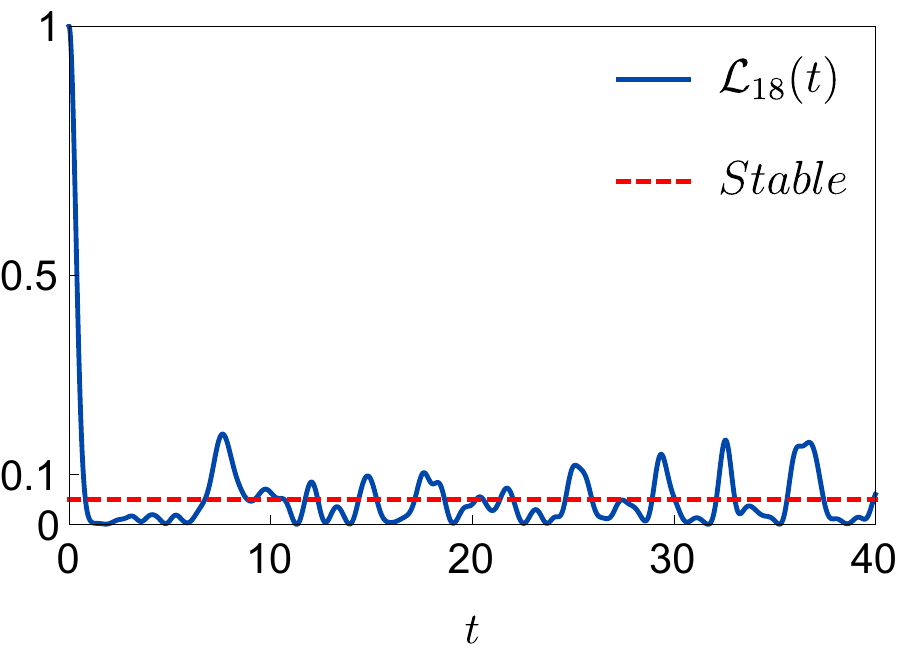}
    \includegraphics[width=0.32\textwidth]{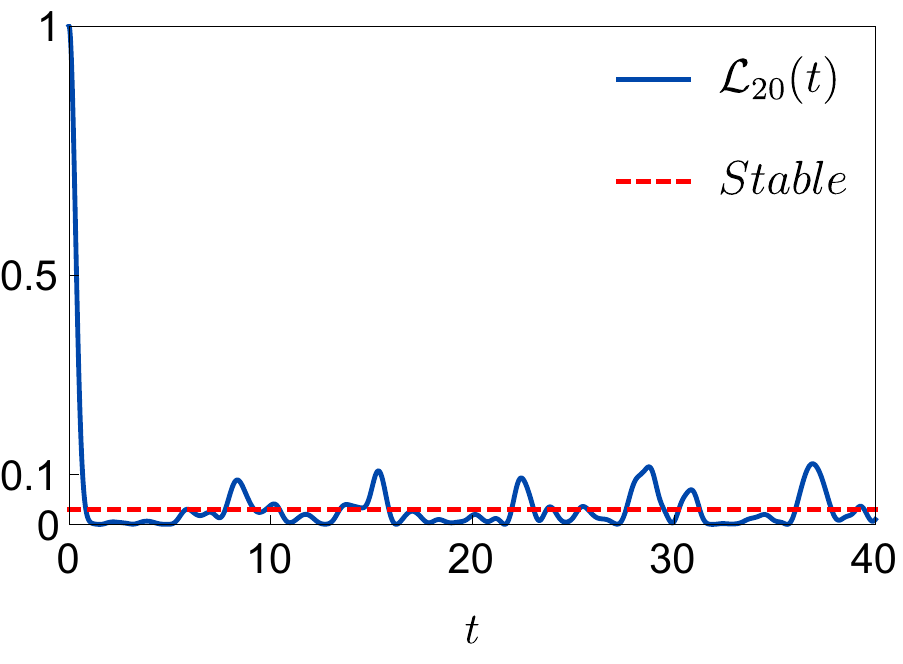}
    \caption{Time evolution of Loschmidt echo for $L=16,18,20$. The blue line are the real time value of the Loschmidt echo and the dashed red line is the predicted stable value, which can also be computed exactly.} 
    \label{Fig.main2} 
\end{figure}

\end{document}